\documentclass[acmtog]{acmart}
\AtBeginDocument{%
  }

\setcopyright{acmlicensed}
\copyrightyear{2025}
\acmYear{2025}
\acmDOI{XXXXXXX.XXXXXXX}

\begin{document}

\title{Principles for Environmental Justice in Technology: Toward a Regenerative Future
}

\author{Sanjana Paul}
\email{sanjana@rootedfutureslab.io}
\orcid{0009-0003-6199-9348}
\affiliation{%
  \institution{Rooted Futures Lab}
  \country{USA}
}
\affiliation{%
  \institution{MIT}
  \country{USA}
}

\renewcommand{\shortauthors}{Paul}

\begin{abstract}
This paper introduces the Environmental Justice in Technology (EJIT) Principles, a framework to help reorient technological development toward social and ecological justice and collective flourishing. In response to prevailing models of technological innovation that prioritize speed, scale, and profit while neglecting systemic injustice, the EJIT principles offer an alternative: a set of guiding values that foreground interdependence, repair, and community self-determination. Drawing inspiration from the 1991 principles of environmental justice, this framework extends their commitments into the technological domain, treating environmental justice not as a peripheral concern but as a necessary foundation for building equitable and regenerative futures. We situate the EJIT principles within the broader landscape of environmental justice, design justice, and post-growth computing, proposing them as a values infrastructure for resisting extractive defaults and envisioning technological systems that operate in reciprocity with people and the planet. In doing so, this article aims to support collective efforts to transform not only what technologies we build, but how, why, and for whom.
\end{abstract}



\received{1 May 2025}
\received[revised]{7 August 2025}
\received[accepted]{10 August 2025}

\maketitle

\section{Introduction}
The accelerating climate crisis and widening social inequities demand a transformation in how technological systems are conceived, built, and governed. Yet dominant models of technology development, driven by speed, scale, and profit, too often reproduce the very extractive logics that fuel environmental degradation and deepen structural injustice. From lithium mining for device batteries to water-intensive data centers and algorithmic systems that entrench surveillance, harm is not an unintended by-product of these systems; it is a feature of their design, deployment, and governance.
\cite{temper_perspective_2018, sharma_post-growth_2024, paul_upgrading_nodate, hine_critical_2023}

Over the last three decades, environmental justice movements, design justice advocates, Indigenous data sovereignty initiatives, and post-growth computing scholarship have all called for centering different aspects of social and environmental justice in technology. \cite{ottinger_environmentally_2012, paul_unsettling_2025, yanchapaxi_indigenous_2025, baker_anti-resilience_nodate} Some of these interventions have exposed how “tech for good” narratives can obscure colonial, capitalist, and militarized entanglements, reducing social and environmental justice to an afterthought or a reputational offset. \cite{paul_upgrading_nodate} While these frameworks have made critical gains, their uptake in practice remains sporadic, often diluted by token consultation, voluntary commitments, or narrow definitions of sustainability that sidestep power and history.

This paper introduces the Environmental Justice in Technology (EJIT) Principles, a collaboratively developed values-based framework inspired by the 1991 Principles of Environmental Justice, which foreground universal protection from harm, self-determination, and the right to a healthy environment. \cite{People_of_Color_Environmental_Leadership_Summit_1991} The EJIT principles extend these commitments into the domain of technological design, development and deployment, based on the belief that technology must not be exempt from the obligations of environmental justice.\cite{van_horne_applied_2023} This framework directly challenges assumptions that technological systems are neutral, inevitable, or inherently beneficial. \cite{malik_critical_2021, birhane_algorithmic_2021, hine_critical_2023} Instead, it insists on centering self-determination of communities, dismantling extractive paradigms, and redefining innovation beyond conventional metrics of success. \cite{paul_upgrading_nodate, temper_blocking_2019}

The EJIT Principles’ distinctive contribution lies in three areas:
\begin{itemize}
\item Explicit anti-neutrality, making clear that every technical decision is political, embedded in histories of extraction or resistance
\item Integration of ecological and socio-political obligations, treating technology as inseparable from the ecosystems and communities it affects
\item Operational scaffolding, organizing principles into thematic clusters that can be adapted for design evaluation, policy assessment, and future innovation. 
\end{itemize}

The EJIT Principles are not a rigid checklist or prescriptive blueprint. Rather, they are a provocation - a set of questions and commitments that reorient the “why” and “how” of technology toward repair, redistribution, and shared governance. Serving as a compass rather than a map, the framework invites researchers, technologists, communities, and policymakers to imagine innovation differently: not as a pursuit of unchecked acceleration, efficiency and profit, but as a practice grounded in justice. In the sections that follow, we situate the EJIT principles within the broader contexts of environmental justice, design justice, and critical innovation studies, articulate the principles themselves, and explore how they might guide the reorientation of technological work toward collective flourishing.

\section{Context and Related Work}
\subsection{Environmental Justice: From Grassroots to Complex Transformation}

The environmental justice movement emerged in the United States in the early 1980s as a grassroots effort driven by communities of color, low-income groups, and Indigenous peoples who organized against the disproportionate siting of environmental hazards in their neighborhoods. \cite{Mohai_Pellow_Roberts_2009} One of the most seminal moments in environmental justice history was the 1982 Warren County, North Carolina protests, where activists mobilized against the placement of a PCB-contaminated landfill in a predominantly Black and low-income community \cite{McGurty_2009}. The protests catalyzed national attention, with major civil rights organizations joining the cause, and led to the US General Accounting Office’s 1983 report documenting the correlation between hazardous waste facilities and the racial composition of surrounding communities(\cite{banzhaf_environmental_2019, agyeman_trends_2016}. Subsequent foundational studies, such as Robert Bullard’s work in Houston \cite{Bullard_1983} and the United Church of Christ’s "Toxic Wastes and Race in the United States" \cite{United_Church_of_Christ_1987}, provided systematic evidence of the direct relationships among pollution, racism, and poverty, demonstrating that communities of color were far more likely to bear the burdens of environmental hazards than their white or more affluent counterparts \cite{banzhaf_environmental_2019, agyeman_trends_2016, TAYLOR_2000}. 

The core concerns of early environmental justice scholarship and activism were focused on these distributional inequalities, which not only exposed glaring omissions in mainstream environmental policy, which were often centered merely on conservation or regulatory compliance, but also recast environmentalism as inseparable from social justice. The movement framed the environment as “where we live, work, and play,” \cite{Novotny_1995} and drew connections to broader struggles for civil rights and equity \cite{vera_when_2019, TAYLOR_2000}. Landmark environmental justice victories, such as President Clinton’s 1994 Executive Order 12898, mandated federal agencies to explicitly address disparate environmental and health impacts on minority and low-income populations, further institutionalizing the demands of grassroots activists within public policy \cite{chowkwanyun_environmental_2023, baker_anti-resilience_nodate, United_Church_of_Christ_1987}.

As environmental justice scholarship and activism developed, its frameworks expanded beyond the borders of the United States, increasingly engaging with global issues such as climate justice, resource extraction, waste colonialism, deforestation, and military testing in Latin America, Africa, Asia, and Indigenous communities facing environmental threats linked to colonial and capitalist expansion \cite{agyeman_trends_2016, ravi_rajan_history_2014, ulloa_perspectives_2017, ogunbode_climate_2022}. Global environmental justice framings bridge local experiences of injustice with transnational dynamics, such as e-waste dumping, global trade agreements, and climate impacts, emphasizing how environmental harms transcend boundaries and frequently amplify systemic and historical inequalities between nations and peoples. \cite{agyeman_trends_2016, ravi_rajan_history_2014, ulloa_perspectives_2017, ogunbode_climate_2022}.

In contemporary scholarship, environmental justice is now recognized as a multidimensional concept that encompasses not only the distribution of environmental burdens and benefits, but also questions of recognition, participation, and epistemic justice. \cite{agyeman_trends_2016, Blue_Bronson_Lajoie-O’Malley_2021, temper_perspective_2018, scognamiglio_bridging_2024}. Multidimensional analysis foregrounds how social, historical, and political context profoundly shape environmental governance, risk assessment, and exclusion. Environmental justice activists and practitioners, as well as scholars like Ottinger and Cohen have shown that marginalized communities are often excluded from decision making, their knowledge undervalued, and their participation tokenized within technocratic or market-driven solutions \cite{ottinger_environmentally_2012}.

Environmental justice as a concept is now distinguished by its plurality, intersectionality, and persistent focus on the underlying structures - racism, capitalism, patriarchy, and colonialism - that produce environmental harm. Global environmental justice scholarship urges that effective remedies require not just distributive fixes, but deep shifts in power, knowledge, and governance practices, supporting grassroots and community-led movements that seek transformative change \cite{temper_blocking_2019, temper_blocking_2019-1, agyeman_trends_2016}. 

\subsection{Indigenous and Plural Epistemologies in Technological Design}

Indigenous and community-rooted frameworks orient the “why” and “how” of technology toward practices not of extraction or exclusion, but of collective care, repair, and shared governance. These epistemologies reject the persistent dualism in mainstream design and environmental policy that separates nature from technology or culture. Instead, they foreground the centrality of relationality: seeing humans, nonhuman beings, and technologies as co-constituted in networks of kinship and mutual responsibility. In this view, technological innovation is inseparable from the social, ecological, and spiritual landscapes in which it is embedded, and ethical action arises from an awareness of reciprocal obligations toward all relations, human and more-than-human alike \cite{deloria_power_2001, lewis_making_2018, escobar_designs_2018, arsenault_including_2019}. 

Such perspectives directly contest dominant narratives of technological “solutionism” and neutrality. By understanding technology as always already entangled with histories of colonialism, capitalism, and state power \cite{baker_anti-resilience_nodate, sayyed_satellite_2024, escobar_designs_2018, ahlborg_bringing_2019, ottinger_environmentally_2012}. Indigenous and plural epistemologies reveal how innovation can compound or repair harm depending on whose knowledge, needs, and values are centered. Relational design practices invite practitioners to move beyond superficial fixes or after-the-fact justice add-ons, insisting instead that genuine accountability, consent, and care must be structured into all stages of technological governance and use \cite{yanchapaxi_indigenous_2025, vera_when_2019, escobar_designs_2018}.

Practically, this means honoring Indigenous sovereignty and governance protocols, and centering collective wisdom and experience as foundational to technology development, not as resources to be extracted or consulted post hoc, but as worldviews with their own intellectual traditions, authorities, and obligations \cite{van_horne_applied_2023, yanchapaxi_indigenous_2025, Blue_Bronson_Lajoie-O’Malley_2021}. Kinship and reciprocity become not just ethical aspirations but operational imperatives: meaningful participation, equitable redistribution of power and benefits, and just processes of repair for past and ongoing harms are all dependent on these relationships \cite{alier_environmentalism_nodate, lewis_making_2018, escobar_designs_2018, davis_participatory_2021}.

\subsection{Beyond Solutionism and Mythologies of Neutrality}
Efforts to build environmentally just futures in technology must contend with the persistent reality that environmental justice is rarely treated as a baseline requirement in the design and deployment of digital and hardware systems. The prominence of terms like ‘tech for good,’  exposes underlying paradigms that are extractive, exclusionary, and complicit in producing or deepening harm, especially for those already made vulnerable. If these frameworks were not needed, it would mean technology and environmental systems inherently benefited marginalized communities and fostered justice and sustainability. Instead, the invocation of justice in tech serves as an indictment of a much larger “tech for bad” world, \cite{paul_upgrading_nodate, 10292177} where care and accountability are exceptions rather than design fundamentals.

A critical strand of scholarship in science and technology studies, design justice, and human-computer interaction challenges the pervasive myth of technological neutrality and "solutionism" in environmental governance. Rather than seeing technology as a detached tool or a neutral apparatus for improvement, scholars like Liboiron, Baker, Malik and Malik, Berg, Ottinger and Cohen, Hine, and others \cite{liboiron_pollution_2021, baker_anti-resilience_nodate, malik_critical_2021, berg_politics_1998, ottinger_environmentally_2012, hine_critical_2023} insist that design, innovation, and environmental systems are always entangled with social, political, and historical power relations. For example, the design justice framework explicitly argues that technological design often reproduces entrenched structures of domination, including white supremacy, capitalism, patriarchy, and settler colonialism, whether or not this reproduction is intentional\cite{costanza-chock_design_nodate}.Technologies, according to Costanza-Chock, are not isolated objects of progress; they “embody social relations (power)” and frequently reflect and reinforce the interests of influential actors while marginalizing others \cite{costanza-chock_design_nodate}.

Similarly, environmental justice scholars highlight how well-intentioned "tech for good" \cite{richey_perils_2024, paul_upgrading_nodate} initiatives can obscure the colonial, capitalist, or militarized entanglements of environmental harm. When environmental justice is approached as an add-on rather than a foundation for design, the result is often superficial fixes that leave deeper inequalities unaddressed\cite{pellow_we_2009, van_de_poel_design_nodate, escobar_designs_2018}. These responses tend to reify existing power structures, hardening the boundaries of who benefits and who bears the burdens, rather than dismantling systems of extraction and exclusion \cite{sikor_globalizing_2014, vera_when_2019, birhane_algorithmic_2021}. As Shalanda Baker notes in her work on energy transitions, resilience-oriented solutions risk “operating to harden existing social, economic, and environmental injustices that disproportionately burden the poor and people of color,” reinforcing the problematic status quo instead of fostering genuine liberation or repair \cite{baker_anti-resilience_nodate}.

The literature also points to the frequent exclusion of those most affected by environmental hazards and technological change \cite{liboiron_pollution_2021, pellow_we_2009, malik_critical_2021, van_horne_applied_2023, banzhaf_environmental_2019}. Innovations, particularly those arising from growth-centric or market-driven logic, habitually prioritize the voices and interests of the powerful: corporations, bureaucratic elites, technologists far removed from the consequences of their designs. Indigenous communities, grassroots organizations, and marginalized populations are often sidelined as passive recipients rather than empowered agents or innovators \cite{tuck_decolonization_nodate, lewis_making_2018, escobar_designs_2018, dombrowski_social_2016, Whyte_2016}.

\section{The EJIT Principles: A Framework for the Future}

The Environmental Justice in Technology (EJIT) Principles distill insights from environmental justice, design justice, post-growth computing, and Indigenous epistemologies into a framework for re-centering technology around repair, reciprocity, and self-determination. Developed through dialogue with communities, technologists, and scholars, and inspired by the 1991 Principles of Environmental Justice, EJIT extends those foundational commitments - universal protection from harm, the right to self-determination, and a healthy environment - into the technological sphere.
Rather than a prescriptive checklist, the EJIT Principles function as provocations: questions and commitments designed to unsettle dominant assumptions about innovation and to shift technological work toward justice, care, and collective flourishing. They ask not only what technologies do, but who they serve, who they harm, and who decides. Organized into four thematic clusters, the principles offer a scaffold for embedding environmental justice into the purpose, process, and governance of technological systems. They are intended to guide researchers, technologists, and communities in evaluating and shaping technological systems through a justice-centered lens.
The EJIT Principles center repair, redistribution, and relational responsibility. They ask not only what technologies do, but who they serve, who they harm, and who decides. They challenge the myth of neutrality and the logic of optimization, proposing instead a regenerative approach to computing and technology development, one that is cognizant of ecological limits and social considerations. The principles are organized into four thematic clusters, each representing a critical shift required to embed environmental justice into technological development.
The principles are as follows:

\begin{itemize}
\item Environmentally just technology is explicitly anti-racist. 
\item Environmental justice in technology calls for responsible innovation in every aspect of technological creation. Responsible innovation occurs when all people are provided the resources to innovate, all potential uses of the innovation are accounted for to prepare for contingencies, emphathy is central to innovation and its creative intent.
\item Environmental justice in technology empowers those who wish to live without certain technologies. It demands preserving traditional Indigenous ways of living without interference from capitalist and corporate technologies. 
\item Environmental justice in tech means refusing to cooperate with or arm the military-industrial complex, prisons, or police. Environmentally just tech is used to elevate all ordinary people, not to oppress any of them with violence or the threat of it. 
\item Environmental justice in technology demands that democracy be the foundation of all of its endeavors. A democratic and community-centric environment is necessary to have a just world. 
\item Environmentally just technology dismantles capitalist-centric development and does not harm economic vitality. It promotes equitable and just income (re)distribution across the world. 
\item Environmentally just technology preserves the beauty and utility of the natural world for future generations. 
\item Environmental justice in technology means having a harmonious relationship with the Earth and with all life. Environmentally just tech has a collaborative, regenerative, and sustainable relationship with the natural world, not an extractive relationship. 
\item Environmentally just technology is not used to exclude parts of the Earth for use by some individuals and not others. It enables all people to access all parts of the Earth. 
\item Environmentally just technology is open-source. Environmentally just tech makes all information about its creation (including blueprints, instructions/manuals, and information for repair) freely available and accessible to empower everyone to make, repair, modify, and develop their technology. 
\item Environmental justice in technology requires that the burdens and benefits of technology be equally shared amongst all people. Environmentally just tech will never empower one group at the expense of another. 
\item Environmental justice in technology strives to eliminate global and local burdens inherent in its creation. 
\item Environmental justice in technology calls for the deployment of technology where and when it is appropriate and beneficial to its local community. When these criteria are not met, environmentally just technology is not deployed. 
\item Environmental justice in technology provides for the cleanup and restoration of lands, waters, and communities that have been harmed by past uses of technology. 
\item Environmental justice in technology calls for the removal of colonial and neocolonial intentions with technology; instead, it encourages self-determination, freedom, and repatriation. 
\item Environmentally just technology is not separate from nature, Earth, and the environment. Rather, it works synergistically with nature. 
\item Environmentally Just Tech is intentional about harm. It is cognizant of who a given technology helps and who it harms.

\end{itemize}

Below, we present the EJIT Principles as a series of prompts and values. Each principle is framed as a question or provocation, inviting reflection and action rather than serving as a rigid checklist. The principles are grouped into four overarching themes, each representing a critical shift required to center environmental justice in technological development.

\subsection{Designing with Power and Positionality in Mind}

There is no such thing as neutral technology. Every technical decision reflects and reinforces structures of power. The EJIT Principles begin by naming this openly: environmentally just technologies must be anti-racist and anti-colonial by design, not just in rhetoric. They must reject relationships with systems of harm that have long relied on technological tools to surveil, discipline, and displace \cite{benjamin2019}, \cite{Noble_2018}.

\begin{itemize}
    \item Principle 1: Is this technology explicitly anti-racist and anti-colonial in its design and intent?
    \item Principle 2: Does this technology empower communities to refuse imposed or extractive technologies?
    \item Principle 3: Who holds decision-making power over this technology? Are frontline communities leading its development and deployment?
\end{itemize}

\subsection{Restructuring Innovation for Collective Flourishing}

Innovation, as it is commonly practiced, rewards speed, novelty, and capital. The EJIT framework pushes for an approach that emphasizes responsibility, empathy, and shared benefit. This includes accounting for unintended consequences, ensuring all people have access to the tools and knowledge needed to shape innovation, and prioritizing collective well-being over market success \cite{paul_upgrading_nodate}.

Innovation here is plural and distributed. It includes Indigenous ecological knowledge, mutual aid infrastructure, grassroots environmental monitoring, and other practices that rarely get recognized in mainstream tech spaces but are no less vital  \cite{escobar_designs_2018}.

\begin{itemize}
    \item Principle 4: Are all people provided with the resources and knowledge to shape innovation?
    \item Principle 5: Are the potential uses and unintended consequences of this technology accounted for and prepared for?
    \item Principle 6: Is empathy central to the technology’s creative intent and impact?
\end{itemize}

\subsection{Reorienting the Relationship Between Technology and Nature}

The idea that technology is somehow outside or above nature is one of the most dangerous assumptions in modern infrastructure. The EJIT Principles treat all technological systems as embedded in ecosystems, materially, energetically, and ethically.

This means prioritizing systems that are regenerative rather than extractive, designed with local ecological limits in mind, and developed in ways that do not worsen environmental degradation. It also means being accountable for harm: not just avoiding future damage, but actively cleaning up and restoring communities and ecosystems harmed by past technologies \cite{clare_assessing_2023}. This draws from a long lineage of work that connects environmental repair with political accountability and ecological humility \cite{tuck_decolonization_nodate}.

\begin{itemize}
    \item Principle 7: Does this technology preserve the beauty and utility of the natural world for future generations?
    \item Principle 8: Does it have a regenerative, not extractive, relationship with nature?
    \item Principle 9: Is it accountable for harm, including the cleanup and restoration of lands, waters, and communities harmed by past technologies?
\end{itemize}

\subsection{Embedding Access, Accountability, and Reparative Practice}

Environmental justice in technology also demands that access and accountability be built into the foundation of a system, not bolted on afterward. This includes open access to documentation, repair tools, and modification rights; it also includes transparency around who benefits, who is excluded, and how harms are addressed \cite{costanza-chock_design_nodate}.

\begin{itemize}
    \item Principle 10: Is all information about this technology’s creation (blueprints, manuals, repair guides) openly accessible?
    \item Principle 11: Are the burdens and benefits of this technology equally shared?
    \item Principle 12: Does this technology enable self-determination, freedom, and repatriation, especially for communities impacted by colonialism?
\end{itemize}

The EJIT Principles do not constitute an exhaustive checklist but function as a conceptual framework of inquiries, provocations, and core values aimed at preventing technological development from perpetuating systemic violence. They provide a structural orientation for establishing repair, redistribution, and relational responsibility as foundational to technological systems, rather than exceptional considerations.

\section{Implications and Applications}

The EJIT Principles are designed to move beyond rhetorical commitments and serve as a practical compass for guiding technology research, design, deployment, and governance. Operationalizing them requires translating their values into clear, measurable criteria, and pairing those criteria with concrete mechanisms that facilitate accountability. Put simply: the EJIT Principles are not only “what to aim for” but “how to do it.”
By drawing on interdisciplinary scholarship, we frame the operationalization of the EJIT Principles as a necessary approach to capturing the complex environmental and justice dimensions of technological innovation. The criteria and mechanisms outlined in the following points serve as a foundational starting point to facilitate ongoing interdisciplinary dialogue and empirical investigation among researchers, policymakers, technologists, and community members. This framework aims to empower diverse actors who are positioned to explore and address the systemic impacts of technology through an integrated lens, grounded in environmental justice scholarship. The questions and criteria below integrate (1) the structural complexities inherent to socio-ecological and socio-technical dynamics, and (2) emergent themes in environmental justice practice and theory.

{\bfseries Power Redistribution:} Does the system transfer decision-making authority to frontline and historically marginalized communities? This echoes calls from both environmental justice and design justice to center meaningful participation and self-determination as foundational to any just outcomes \cite{davis_participatory_2021}.

{\bfseries Harm Repair:} Does the project include explicit plans for restoring ecosystems and communities affected by past technological harms? This reflects a core aspect of the environmental justice movement’s emphasis on accountability and reparative justice \cite{chowkwanyun_environmental_2023}. 

{\bfseries Careful Open Access:} Are blueprints, repair guides, and data publicly accessible? Open-source and open-access practices can be a key part of efforts to democratize technology and enable community-led innovation, where appropriate \cite{sharma_post-growth_2024}.

{\bfseries Contextual Deployment:} Is technology deployed only where it is locally beneficial and appropriate? Paying attention to this criterion ensures that technological interventions are responsive to community needs and ecological limits \cite{scognamiglio_bridging_2024}.

These criteria could inform funding, policy, and design benchmarks, such as requiring community consent and reparative commitments in grant applications or technology procurement processes.

\subsection{From Principles to Practice}

Translating the EJIT Principles from conceptual commitments into tangible impact requires embedding them across multiple dimensions of technological work. By operationalizing justice-centered values in education, project evaluation, development approaches, and infrastructure governance, the principles can effectively influence the culture, processes, and outcomes of technology innovation. This multi-pronged approach fosters systemic change by cultivating knowledge, shaping incentives, and empowering communities.

{\bfseries Integrating environmental justice:} Integrating environmental justice as a core concern within computing, engineering, and design curricula is a foundational step for cultivating a new generation of technologists equipped to understand and address complex social-ecological challenges. Embedding justice-oriented frameworks and participatory methods encourages critical reflection on the societal and environmental implications of technology from the outset of training. Such educational reform aligns with recent scholarship advocating for interdisciplinary pedagogy that merges ethical reasoning, community engagement, and technical competence to produce justice-attuned innovation leaders \cite{van_de_poel_design_nodate, paul_unsettling_2025}.

{\bfseries Project Assessment and Evaluation:} Moving beyond traditional metrics focused narrowly on usability, efficiency, or market viability, project assessments must integrate multidimensional justice criteria that examine distribution of benefits and burdens, recognition of diverse knowledge systems, and meaningful participation of affected communities. Assessments informed by this expanded justice lens, drawing on emerging frameworks from environmental justice, design justice, and critical innovation studies, can better capture the socio-ecological impacts of technology and support decisions that prioritize equitable and reparative outcomes \cite{scognamiglio_bridging_2024, massarella_reproducing_2020, Blue_Bronson_Lajoie-O’Malley_2021}.

{\bfseries Open-Source and Community-Led Development:}  Prioritizing open-source principles that emphasize transparency, repairability, and community governance cultivates more equitable technological ecosystems. Such approaches empower users and communities to modify, maintain, and adapt technologies according to local needs and values, facilitating distributed innovation and shared stewardship. By centering community leadership and governance in technology development, the benefits and responsibilities of innovation can be more fairly distributed, while also enhancing resilience and \cite{van_de_poel_design_nodate, sharma_post-growth_2024}.

\section{Conclusion}

To compute within limits is not merely to constrain, but to fundamentally reimagine the purpose, process, and politics of technological creation. It requires us to critically interrogate what should be built, who holds the authority to make such decisions, and whether construction itself is always the most ethical or necessary response. The EJIT Principles provide a conceptual map for this reorientation: a framework of values, provocations, and practices designed to guide us through the complex intersections of planetary crisis, technological overreach, and collective survival.
These principles compel us to move beyond inherited measures of success, such as scale, speed, and profitability, and to adopt alternative metrics rooted in repair, redistribution, and regeneration. They demand that we confront the systemic harms encoded within computing’s supply chains, funding structures, and imagined users, and that we commit to building technological systems that are accountable to both people and place. Recent scholarship on environmental data justice and community-based participatory research underscores the importance of such accountability, emphasizing the need to center historically marginalized voices and to challenge extractive logics in technological design. 
The work of justice-oriented technology is neither neutral nor optional. As the climate crisis intensifies and extractive technologies proliferate, computing must become a site of principled refusal, a space where business-as-usual is actively resisted and alternative futures are deliberately designed. The EJIT Principles offer a starting point for this transformation, grounding design in justice, amplifying the perspectives of those who have long resisted environmental harm, and reminding us that technology is never exempt from the ethical and ecological obligations of our time. We offer this framework not as a definitive solution, but as an open invitation: to collaboratively construct new models of environmental innovation, to center justice in every technical choice, and to approach every byte stream, build, and blueprint as an opportunity to enact care. In doing so, we acknowledge the ongoing work of environmental justice communities, the necessity of continuous self-reflection and accountability, and the imperative to build technological futures that are equitable, regenerative, and just.

\citestyle{acmnumeric}    
\bibliographystyle{ACM-Reference-Format}
\bibliography{Limits2025}


\begin{thebibliography}{52}


\ifx \showCODEN    \undefined \def \showCODEN     #1{\unskip}     \fi
\ifx \showISBNx    \undefined \def \showISBNx     #1{\unskip}     \fi
\ifx \showISBNxiii \undefined \def \showISBNxiii  #1{\unskip}     \fi
\ifx \showISSN     \undefined \def \showISSN      #1{\unskip}     \fi
\ifx \showLCCN     \undefined \def \showLCCN      #1{\unskip}     \fi
\ifx \shownote     \undefined \def \shownote      #1{#1}          \fi
\ifx \showarticletitle \undefined \def \showarticletitle #1{#1}   \fi
\ifx \showURL      \undefined \def \showURL       {\relax}        \fi
\providecommand\bibfield[2]{#2}
\providecommand\bibinfo[2]{#2}
\providecommand\natexlab[1]{#1}
\providecommand\showeprint[2][]{arXiv:#2}

\bibitem[Agyeman et~al\mbox{.}(2016)]%
        {agyeman_trends_2016}
\bibfield{author}{\bibinfo{person}{Julian Agyeman}, \bibinfo{person}{David Schlosberg}, \bibinfo{person}{Luke Craven}, {and} \bibinfo{person}{Caitlin Matthews}.} \bibinfo{year}{2016}\natexlab{}.
\newblock \showarticletitle{Trends and {Directions} in {Environmental} {Justice}: {From} {Inequity} to {Everyday} {Life}, {Community}, and {Just} {Sustainabilities}}.
\newblock \bibinfo{journal}{\emph{Annual Review of Environment and Resources}} \bibinfo{volume}{41}, \bibinfo{number}{1} (\bibinfo{date}{Nov.} \bibinfo{year}{2016}), \bibinfo{pages}{321--340}.
\newblock
\showISSN{1543-5938, 1545-2050}
\href{https://doi.org/10.1146/annurev-environ-110615-090052}{doi:\nolinkurl{10.1146/annurev-environ-110615-090052}}


\bibitem[Ahlborg et~al\mbox{.}(2019)]%
        {ahlborg_bringing_2019}
\bibfield{author}{\bibinfo{person}{Helene Ahlborg}, \bibinfo{person}{Ilse Ruiz-Mercado}, \bibinfo{person}{Sverker Molander}, {and} \bibinfo{person}{Omar Masera}.} \bibinfo{year}{2019}\natexlab{}.
\newblock \showarticletitle{Bringing {Technology} into {Social}-{Ecological} {Systems} {Research}—{Motivations} for a {Socio}-{Technical}-{Ecological} {Systems} {Approach}}.
\newblock \bibinfo{journal}{\emph{Sustainability}} \bibinfo{volume}{11}, \bibinfo{number}{7} (\bibinfo{date}{April} \bibinfo{year}{2019}), \bibinfo{pages}{2009}.
\newblock
\showISSN{2071-1050}
\href{https://doi.org/10.3390/su11072009}{doi:\nolinkurl{10.3390/su11072009}}


\bibitem[Alier(2002)]%
        {alier_environmentalism_nodate}
\bibfield{author}{\bibinfo{person}{Joan~Martinez Alier}.} \bibinfo{year}{2002}\natexlab{}.
\newblock \showarticletitle{The {Environmentalism} of the {Poor}}.
\newblock  (\bibinfo{year}{2002}).
\newblock


\bibitem[Arsenault et~al\mbox{.}(2019)]%
        {arsenault_including_2019}
\bibfield{author}{\bibinfo{person}{Rachel Arsenault}, \bibinfo{person}{Carrie Bourassa}, \bibinfo{person}{Sibyl Diver}, \bibinfo{person}{Deborah McGregor}, {and} \bibinfo{person}{Aaron Witham}.} \bibinfo{year}{2019}\natexlab{}.
\newblock \showarticletitle{Including {Indigenous} {Knowledge} {Systems} in {Environmental} {Assessments}: {Restructuring} the {Process}}.
\newblock \bibinfo{journal}{\emph{Global Environmental Politics}} \bibinfo{volume}{19}, \bibinfo{number}{3} (\bibinfo{date}{Aug.} \bibinfo{year}{2019}), \bibinfo{pages}{120--132}.
\newblock
\showISSN{1526-3800, 1536-0091}
\href{https://doi.org/10.1162/glep_a_00519}{doi:\nolinkurl{10.1162/glep_a_00519}}


\bibitem[Baker({[n.\,d.]})]%
        {baker_anti-resilience_nodate}
\bibfield{author}{\bibinfo{person}{Shalanda~H Baker}.} \bibinfo{year}{[n.\,d.]}\natexlab{}.
\newblock \showarticletitle{Anti-{Resilience}: {A} {Roadmap} for {Transformational} {Justice} within the {Energy} {System}}.
\newblock \bibinfo{journal}{\emph{Harvard Civil Rights}}  \bibinfo{volume}{54} (\bibinfo{year}{[n.\,d.]}).
\newblock


\bibitem[Banzhaf et~al\mbox{.}(2019)]%
        {banzhaf_environmental_2019}
\bibfield{author}{\bibinfo{person}{H.~Spencer Banzhaf}, \bibinfo{person}{Lala Ma}, {and} \bibinfo{person}{Christopher Timmins}.} \bibinfo{year}{2019}\natexlab{}.
\newblock \showarticletitle{Environmental {Justice}: {Establishing} {Causal} {Relationships}}.
\newblock \bibinfo{journal}{\emph{Annual Review of Resource Economics}} \bibinfo{volume}{11}, \bibinfo{number}{1} (\bibinfo{date}{Oct.} \bibinfo{year}{2019}), \bibinfo{pages}{377--398}.
\newblock
\showISSN{1941-1340, 1941-1359}
\href{https://doi.org/10.1146/annurev-resource-100518-094131}{doi:\nolinkurl{10.1146/annurev-resource-100518-094131}}


\bibitem[Benjamin(2019)]%
        {benjamin2019}
\bibfield{author}{\bibinfo{person}{Ruha Benjamin}.} \bibinfo{year}{2019}\natexlab{}.
\newblock \bibinfo{booktitle}{\emph{Race After Technology: Abolitionist Tools for the New Jim Code}}.
\newblock \bibinfo{publisher}{Polity}, \bibinfo{address}{Cambridge, UK; Medford, MA}.
\newblock


\bibitem[Berg(1998)]%
        {berg_politics_1998}
\bibfield{author}{\bibinfo{person}{Marc Berg}.} \bibinfo{year}{1998}\natexlab{}.
\newblock \showarticletitle{The {Politics} of {Technology}: {On} {Bringing} {Social} {Theory} into {Technological} {Design}}.
\newblock \bibinfo{journal}{\emph{Science, Technology, \& Human Values}} \bibinfo{volume}{23}, \bibinfo{number}{4} (\bibinfo{date}{Oct.} \bibinfo{year}{1998}), \bibinfo{pages}{456--490}.
\newblock
\showISSN{0162-2439, 1552-8251}
\href{https://doi.org/10.1177/016224399802300406}{doi:\nolinkurl{10.1177/016224399802300406}}


\bibitem[Birhane(2021)]%
        {birhane_algorithmic_2021}
\bibfield{author}{\bibinfo{person}{Abeba Birhane}.} \bibinfo{year}{2021}\natexlab{}.
\newblock \showarticletitle{Algorithmic injustice: a relational ethics approach}.
\newblock \bibinfo{journal}{\emph{Patterns}} \bibinfo{volume}{2}, \bibinfo{number}{2} (\bibinfo{date}{Feb.} \bibinfo{year}{2021}), \bibinfo{pages}{100205}.
\newblock
\showISSN{26663899}
\href{https://doi.org/10.1016/j.patter.2021.100205}{doi:\nolinkurl{10.1016/j.patter.2021.100205}}


\bibitem[Blue et~al\mbox{.}(2021)]%
        {Blue_Bronson_Lajoie-O’Malley_2021}
\bibfield{author}{\bibinfo{person}{Gwendolyn Blue}, \bibinfo{person}{Kelly Bronson}, {and} \bibinfo{person}{Alana Lajoie-O’Malley}.} \bibinfo{year}{2021}\natexlab{}.
\newblock \showarticletitle{Beyond distribution and participation: A scoping review to advance a comprehensive environmental justice framework for impact assessment}.
\newblock \bibinfo{journal}{\emph{Environmental Impact Assessment Review}}  \bibinfo{volume}{90} (\bibinfo{date}{Sep} \bibinfo{year}{2021}), \bibinfo{pages}{106607}.
\newblock
\href{https://doi.org/10.1016/j.eiar.2021.106607}{doi:\nolinkurl{10.1016/j.eiar.2021.106607}}


\bibitem[Brooks et~al\mbox{.}(2023)]%
        {10292177}
\bibfield{author}{\bibinfo{person}{Ian Brooks}, \bibinfo{person}{Minna~Laurell Thorslund}, {and} \bibinfo{person}{Aksel Bi¢rn-Hansen}.} \bibinfo{year}{2023}\natexlab{}.
\newblock \showarticletitle{Tech4Bad in the Oil and Gas Industry: Exploring Choices for ICT Professionals}. In \bibinfo{booktitle}{\emph{2023 International Conference on ICT for Sustainability (ICT4S)}}. \bibinfo{pages}{142--153}.
\newblock
\href{https://doi.org/10.1109/ICT4S58814.2023.00023}{doi:\nolinkurl{10.1109/ICT4S58814.2023.00023}}


\bibitem[Bullard(1983)]%
        {Bullard_1983}
\bibfield{author}{\bibinfo{person}{Robert~D. Bullard}.} \bibinfo{year}{1983}\natexlab{}.
\newblock \showarticletitle{Solid waste sites and the Black Houston Community*}.
\newblock \bibinfo{journal}{\emph{Sociological Inquiry}} \bibinfo{volume}{53}, \bibinfo{number}{2–3} (\bibinfo{date}{Apr} \bibinfo{year}{1983}), \bibinfo{pages}{273–288}.
\newblock
\href{https://doi.org/10.1111/j.1475-682x.1983.tb00037.x}{doi:\nolinkurl{10.1111/j.1475-682x.1983.tb00037.x}}


\bibitem[Chowkwanyun(2023)]%
        {chowkwanyun_environmental_2023}
\bibfield{author}{\bibinfo{person}{Merlin Chowkwanyun}.} \bibinfo{year}{2023}\natexlab{}.
\newblock \showarticletitle{Environmental {Justice}: {Where} {It} {Has} {Been}, and {Where} {It} {Might} {Be} {Going}}.
\newblock \bibinfo{journal}{\emph{Annual Review of Public Health}} \bibinfo{volume}{44}, \bibinfo{number}{1} (\bibinfo{date}{April} \bibinfo{year}{2023}), \bibinfo{pages}{93--111}.
\newblock
\showISSN{0163-7525, 1545-2093}
\href{https://doi.org/10.1146/annurev-publhealth-071621-064925}{doi:\nolinkurl{10.1146/annurev-publhealth-071621-064925}}


\bibitem[Clare et~al\mbox{.}(2023)]%
        {clare_assessing_2023}
\bibfield{author}{\bibinfo{person}{M.~A. Clare}, \bibinfo{person}{A. Lichtschlag}, \bibinfo{person}{S. Paradis}, {and} \bibinfo{person}{N.~L.~M. Barlow}.} \bibinfo{year}{2023}\natexlab{}.
\newblock \showarticletitle{Assessing the impact of the global subsea telecommunications network on sedimentary organic carbon stocks}.
\newblock \bibinfo{journal}{\emph{Nature Communications}} \bibinfo{volume}{14}, \bibinfo{number}{1} (\bibinfo{date}{April} \bibinfo{year}{2023}), \bibinfo{pages}{2080}.
\newblock
\showISSN{2041-1723}
\href{https://doi.org/10.1038/s41467-023-37854-6}{doi:\nolinkurl{10.1038/s41467-023-37854-6}}


\bibitem[Costanza-Chock(2020)]%
        {costanza-chock_design_nodate}
\bibfield{author}{\bibinfo{person}{Sasha Costanza-Chock}.} \bibinfo{year}{2020}\natexlab{}.
\newblock \showarticletitle{Design {Justice}: {Community}-{Led} {Practices} to {Build} the {Worlds} {We} {Need}}.
\newblock  (\bibinfo{year}{2020}).
\newblock


\bibitem[Davis and Ramírez-Andreotta(2021)]%
        {davis_participatory_2021}
\bibfield{author}{\bibinfo{person}{Leona~F. Davis} {and} \bibinfo{person}{Mónica~D. Ramírez-Andreotta}.} \bibinfo{year}{2021}\natexlab{}.
\newblock \showarticletitle{Participatory {Research} for {Environmental} {Justice}: {A} {Critical} {Interpretive} {Synthesis}}.
\newblock \bibinfo{journal}{\emph{Environmental Health Perspectives}} \bibinfo{volume}{129}, \bibinfo{number}{2} (\bibinfo{date}{Feb.} \bibinfo{year}{2021}), \bibinfo{pages}{026001}.
\newblock
\showISSN{0091-6765, 1552-9924}
\href{https://doi.org/10.1289/EHP6274}{doi:\nolinkurl{10.1289/EHP6274}}


\bibitem[Deloria and Wildcat(2001)]%
        {deloria_power_2001}
\bibfield{author}{\bibinfo{person}{Vine Deloria} {and} \bibinfo{person}{Daniel~R. Wildcat}.} \bibinfo{year}{2001}\natexlab{}.
\newblock \bibinfo{booktitle}{\emph{Power and place: {Indian} education in {America}}}.
\newblock \bibinfo{publisher}{Fulcrum}, \bibinfo{address}{Golden (Colo.)}.
\newblock
\showISBNx{978-1-55591-859-0}


\bibitem[Dombrowski et~al\mbox{.}(2016)]%
        {dombrowski_social_2016}
\bibfield{author}{\bibinfo{person}{Lynn Dombrowski}, \bibinfo{person}{Ellie Harmon}, {and} \bibinfo{person}{Sarah Fox}.} \bibinfo{year}{2016}\natexlab{}.
\newblock \showarticletitle{Social {Justice}-{Oriented} {Interaction} {Design}: {Outlining} {Key} {Design} {Strategies} and {Commitments}}. In \bibinfo{booktitle}{\emph{Proceedings of the 2016 {ACM} {Conference} on {Designing} {Interactive} {Systems}}}. \bibinfo{publisher}{ACM}, \bibinfo{address}{Brisbane QLD Australia}, \bibinfo{pages}{656--671}.
\newblock
\showISBNx{978-1-4503-4031-1}
\href{https://doi.org/10.1145/2901790.2901861}{doi:\nolinkurl{10.1145/2901790.2901861}}


\bibitem[Escobar(2018)]%
        {escobar_designs_2018}
\bibfield{author}{\bibinfo{person}{Arturo Escobar}.} \bibinfo{year}{2018}\natexlab{}.
\newblock \bibinfo{booktitle}{\emph{Designs for the pluriverse: radical interdependence, autonomy, and the making of worlds}}.
\newblock \bibinfo{publisher}{Duke University press}, \bibinfo{address}{Durham}.
\newblock
\showISBNx{978-0-8223-7090-1 978-0-8223-7105-2}


\bibitem[Hine et~al\mbox{.}(2023)]%
        {hine_critical_2023}
\bibfield{author}{\bibinfo{person}{Amelia Hine}, \bibinfo{person}{Chris Gibson}, {and} \bibinfo{person}{Robyn Mayes}.} \bibinfo{year}{2023}\natexlab{}.
\newblock \showarticletitle{Critical minerals: rethinking extractivism?}
\newblock \bibinfo{journal}{\emph{Australian Geographer}} \bibinfo{volume}{54}, \bibinfo{number}{3} (\bibinfo{date}{July} \bibinfo{year}{2023}), \bibinfo{pages}{233--250}.
\newblock
\showISSN{0004-9182, 1465-3311}
\href{https://doi.org/10.1080/00049182.2023.2210733}{doi:\nolinkurl{10.1080/00049182.2023.2210733}}


\bibitem[Lewis et~al\mbox{.}(2018)]%
        {lewis_making_2018}
\bibfield{author}{\bibinfo{person}{Jason~Edward Lewis}, \bibinfo{person}{Noelani Arista}, \bibinfo{person}{Archer Pechawis}, {and} \bibinfo{person}{Suzanne Kite}.} \bibinfo{year}{2018}\natexlab{}.
\newblock \showarticletitle{Making {Kin} with the {Machines}}.
\newblock \bibinfo{journal}{\emph{Journal of Design and Science}} (\bibinfo{date}{July} \bibinfo{year}{2018}).
\newblock
\href{https://doi.org/10.21428/bfafd97b}{doi:\nolinkurl{10.21428/bfafd97b}}


\bibitem[Liboiron(2021)]%
        {liboiron_pollution_2021}
\bibfield{author}{\bibinfo{person}{Max Liboiron}.} \bibinfo{year}{2021}\natexlab{}.
\newblock \bibinfo{booktitle}{\emph{Pollution {Is} {Colonialism}}}.
\newblock \bibinfo{publisher}{Duke University Press}, \bibinfo{address}{Durham London}.
\newblock
\showISBNx{978-1-4780-1322-8 978-1-4780-2144-5}


\bibitem[Malik and Malik(2021)]%
        {malik_critical_2021}
\bibfield{author}{\bibinfo{person}{Maya Malik} {and} \bibinfo{person}{Momin~M. Malik}.} \bibinfo{year}{2021}\natexlab{}.
\newblock \showarticletitle{Critical {Technical} {Awakenings}}.
\newblock \bibinfo{journal}{\emph{Journal of Social Computing}} \bibinfo{volume}{2}, \bibinfo{number}{4} (\bibinfo{date}{Dec.} \bibinfo{year}{2021}), \bibinfo{pages}{365--384}.
\newblock
\showISSN{2688-5255}
\href{https://doi.org/10.23919/JSC.2021.0035}{doi:\nolinkurl{10.23919/JSC.2021.0035}}


\bibitem[Massarella et~al\mbox{.}(2020)]%
        {massarella_reproducing_2020}
\bibfield{author}{\bibinfo{person}{Kate Massarella}, \bibinfo{person}{Susannah~M. Sallu}, {and} \bibinfo{person}{Jonathan~E. Ensor}.} \bibinfo{year}{2020}\natexlab{}.
\newblock \showarticletitle{Reproducing injustice: {Why} recognition matters in conservation project evaluation}.
\newblock \bibinfo{journal}{\emph{Global Environmental Change}}  \bibinfo{volume}{65} (\bibinfo{date}{Nov.} \bibinfo{year}{2020}), \bibinfo{pages}{102181}.
\newblock
\showISSN{09593780}
\href{https://doi.org/10.1016/j.gloenvcha.2020.102181}{doi:\nolinkurl{10.1016/j.gloenvcha.2020.102181}}


\bibitem[McGurty(2009)]%
        {McGurty_2009}
\bibfield{author}{\bibinfo{person}{Eileen~Maura McGurty}.} \bibinfo{year}{2009}\natexlab{}.
\newblock \bibinfo{booktitle}{\emph{Transforming environmentalism: Warren County, PCBS, and the origins of environmental justice}}.
\newblock \bibinfo{publisher}{Rutgers University Press}.
\newblock


\bibitem[Mohai et~al\mbox{.}(2009)]%
        {Mohai_Pellow_Roberts_2009}
\bibfield{author}{\bibinfo{person}{Paul Mohai}, \bibinfo{person}{David Pellow}, {and} \bibinfo{person}{J.~Timmons Roberts}.} \bibinfo{year}{2009}\natexlab{}.
\newblock \showarticletitle{Environmental justice}.
\newblock \bibinfo{journal}{\emph{Annual Review of Environment and Resources}} \bibinfo{volume}{34}, \bibinfo{number}{1} (\bibinfo{date}{Nov} \bibinfo{year}{2009}), \bibinfo{pages}{405–430}.
\newblock
\href{https://doi.org/10.1146/annurev-environ-082508-094348}{doi:\nolinkurl{10.1146/annurev-environ-082508-094348}}


\bibitem[Noble(2018)]%
        {Noble_2018}
\bibfield{author}{\bibinfo{person}{Safiya~Umoja Noble}.} \bibinfo{year}{2018}\natexlab{}.
\newblock \bibinfo{booktitle}{\emph{Algorithms of oppression: How search engines reinforce racism}}.
\newblock \bibinfo{publisher}{New York University Press}.
\newblock


\bibitem[Novotny(1995)]%
        {Novotny_1995}
\bibfield{author}{\bibinfo{person}{Patrick Novotny}.} \bibinfo{year}{1995}\natexlab{}.
\newblock \showarticletitle{Where we live, work and play}.
\newblock \bibinfo{journal}{\emph{New Political Science}} \bibinfo{volume}{16}, \bibinfo{number}{1} (\bibinfo{year}{1995}), \bibinfo{pages}{61–79}.
\newblock
\href{https://doi.org/10.1080/07393149508429738}{doi:\nolinkurl{10.1080/07393149508429738}}


\bibitem[of~Color Environmental Leadership~Summit(1991)]%
        {People_of_Color_Environmental_Leadership_Summit_1991}
\bibfield{author}{\bibinfo{person}{People of Color Environmental Leadership~Summit}.} \bibinfo{year}{1991}\natexlab{}.
\newblock \bibinfo{title}{The principles of Environmental Justice (EJ)}.
\newblock
\urldef\tempurl%
\url{https://www.ejnet.org/ej/principles.pdf}
\showURL{%
\tempurl}


\bibitem[Ogunbode(2022)]%
        {ogunbode_climate_2022}
\bibfield{author}{\bibinfo{person}{Charles~A. Ogunbode}.} \bibinfo{year}{2022}\natexlab{}.
\newblock \showarticletitle{Climate justice is social justice in the {Global} {South}}.
\newblock \bibinfo{journal}{\emph{Nature Human Behaviour}} \bibinfo{volume}{6}, \bibinfo{number}{11} (\bibinfo{date}{Nov.} \bibinfo{year}{2022}), \bibinfo{pages}{1443--1443}.
\newblock
\showISSN{2397-3374}
\href{https://doi.org/10.1038/s41562-022-01456-x}{doi:\nolinkurl{10.1038/s41562-022-01456-x}}


\bibitem[Ottinger and Cohen(2012)]%
        {ottinger_environmentally_2012}
\bibfield{author}{\bibinfo{person}{Gwen Ottinger} {and} \bibinfo{person}{Benjamin Cohen}.} \bibinfo{year}{2012}\natexlab{}.
\newblock \showarticletitle{Environmentally {Just} {Transformations} of {Expert} {Cultures}: {Toward} the {Theory} and {Practice} of a {Renewed} {Science} and {Engineering}}.
\newblock \bibinfo{journal}{\emph{Environmental Justice}} \bibinfo{volume}{5}, \bibinfo{number}{3} (\bibinfo{date}{June} \bibinfo{year}{2012}), \bibinfo{pages}{158--163}.
\newblock
\showISSN{1939-4071, 1937-5174}
\href{https://doi.org/10.1089/env.2010.0032}{doi:\nolinkurl{10.1089/env.2010.0032}}


\bibitem[Paul and Minns(2024)]%
        {paul_upgrading_nodate}
\bibfield{author}{\bibinfo{person}{Sanjana Paul} {and} \bibinfo{person}{Camille Minns}.} \bibinfo{year}{2024}\natexlab{}.
\newblock \showarticletitle{Upgrading {Environmental} {Innovation} for the 21st {Century}}.
\newblock  (\bibinfo{year}{2024}).
\newblock


\bibitem[Paul et~al\mbox{.}(2025)]%
        {paul_unsettling_2025}
\bibfield{author}{\bibinfo{person}{Sanjana Paul}, \bibinfo{person}{Christopher Rabe}, {and} \bibinfo{person}{Camille Minns}.} \bibinfo{year}{2025}\natexlab{}.
\newblock \showarticletitle{Unsettling the {Status} {Quo}: {Embedding} {Environmental} {Justice} in {Tech}-{Centered} {Environmental} {Education}}.
\newblock \bibinfo{journal}{\emph{Contingencies}} \bibinfo{volume}{3}, \bibinfo{number}{1} (\bibinfo{year}{2025}), \bibinfo{pages}{1}.
\newblock
\showISSN{2689-6311}
\href{https://doi.org/10.33682/m5rv-sg6e}{doi:\nolinkurl{10.33682/m5rv-sg6e}}


\bibitem[Pellow(2009)]%
        {pellow_we_2009}
\bibfield{author}{\bibinfo{person}{David~Naguib Pellow}.} \bibinfo{year}{2009}\natexlab{}.
\newblock \showarticletitle{“{We} {Didn}'t {Get} the {First} 500 {Years} {Right}, {So} {Let}'s {Work} on the {Next} 500 {Years}”: {A} {Call} for {Transformative} {Analysis} and {Action}}.
\newblock \bibinfo{journal}{\emph{Environmental Justice}} \bibinfo{volume}{2}, \bibinfo{number}{1} (\bibinfo{date}{March} \bibinfo{year}{2009}), \bibinfo{pages}{3--6}.
\newblock
\showISSN{1939-4071, 1937-5174}
\href{https://doi.org/10.1089/env.2008.0549}{doi:\nolinkurl{10.1089/env.2008.0549}}


\bibitem[Ravi~Rajan(2014)]%
        {ravi_rajan_history_2014}
\bibfield{author}{\bibinfo{person}{S. Ravi~Rajan}.} \bibinfo{year}{2014}\natexlab{}.
\newblock \showarticletitle{A {History} of {Environmental} {Justice} in {India}}.
\newblock \bibinfo{journal}{\emph{Environmental Justice}} \bibinfo{volume}{7}, \bibinfo{number}{5} (\bibinfo{date}{Oct.} \bibinfo{year}{2014}), \bibinfo{pages}{117--121}.
\newblock
\showISSN{1939-4071, 1937-5174}
\href{https://doi.org/10.1089/env.2014.7501}{doi:\nolinkurl{10.1089/env.2014.7501}}


\bibitem[Richey and Fejerskov(2024)]%
        {richey_perils_2024}
\bibfield{author}{\bibinfo{person}{Lisa~Ann Richey} {and} \bibinfo{person}{Adam~Moe Fejerskov}.} \bibinfo{year}{2024}\natexlab{}.
\newblock \showarticletitle{The perils of ‘tech for good’ lie in its politics of helping}.
\newblock \bibinfo{journal}{\emph{Big Data \& Society}} \bibinfo{volume}{11}, \bibinfo{number}{4} (\bibinfo{date}{Dec.} \bibinfo{year}{2024}), \bibinfo{pages}{20539517241267718}.
\newblock
\showISSN{2053-9517, 2053-9517}
\href{https://doi.org/10.1177/20539517241267718}{doi:\nolinkurl{10.1177/20539517241267718}}


\bibitem[Sayyed et~al\mbox{.}(2024)]%
        {sayyed_satellite_2024}
\bibfield{author}{\bibinfo{person}{Tanya~Kreutzer Sayyed}, \bibinfo{person}{Ufuoma Ovienmhada}, \bibinfo{person}{Mitra Kashani}, \bibinfo{person}{Karn Vohra}, \bibinfo{person}{Gaige~Hunter Kerr}, \bibinfo{person}{Catherine O’Donnell}, \bibinfo{person}{Maria~H Harris}, \bibinfo{person}{Laura Gladson}, \bibinfo{person}{Andrea~R Titus}, \bibinfo{person}{Susana~B Adamo}, \bibinfo{person}{Kelvin~C Fong}, \bibinfo{person}{Emily~M Gargulinski}, \bibinfo{person}{Amber~J Soja}, \bibinfo{person}{Susan Anenberg}, {and} \bibinfo{person}{Yusuke Kuwayama}.} \bibinfo{year}{2024}\natexlab{}.
\newblock \showarticletitle{Satellite data for environmental justice: a scoping review of the literature in the {United} {States}}.
\newblock \bibinfo{journal}{\emph{Environmental Research Letters}} \bibinfo{volume}{19}, \bibinfo{number}{3} (\bibinfo{date}{March} \bibinfo{year}{2024}), \bibinfo{pages}{033001}.
\newblock
\showISSN{1748-9326}
\href{https://doi.org/10.1088/1748-9326/ad1fa4}{doi:\nolinkurl{10.1088/1748-9326/ad1fa4}}


\bibitem[Scognamiglio(2024)]%
        {scognamiglio_bridging_2024}
\bibfield{author}{\bibinfo{person}{Giorgia Scognamiglio}.} \bibinfo{year}{2024}\natexlab{}.
\newblock \showarticletitle{{BRIDGING} {THEORETICAL} {ADVANCEMENTS} {AND} {EMPIRICAL} {PRACTICES} {IN} {ENVIRONMENTAL} {JUSTICE} {RESEARCH}: {TOWARDS} {A} {MIXED}-{METHODS} {APPROACH}}.
\newblock  (\bibinfo{year}{2024}).
\newblock


\bibitem[Sharma et~al\mbox{.}(2024)]%
        {sharma_post-growth_2024}
\bibfield{author}{\bibinfo{person}{Vishal Sharma}, \bibinfo{person}{Neha Kumar}, {and} \bibinfo{person}{Bonnie Nardi}.} \bibinfo{year}{2024}\natexlab{}.
\newblock \showarticletitle{Post-growth {Human}–{Computer} {Interaction}}.
\newblock \bibinfo{journal}{\emph{ACM Transactions on Computer-Human Interaction}} \bibinfo{volume}{31}, \bibinfo{number}{1} (\bibinfo{date}{Feb.} \bibinfo{year}{2024}), \bibinfo{pages}{1--37}.
\newblock
\showISSN{1073-0516, 1557-7325}
\href{https://doi.org/10.1145/3624981}{doi:\nolinkurl{10.1145/3624981}}


\bibitem[Sikor and Newell(2014)]%
        {sikor_globalizing_2014}
\bibfield{author}{\bibinfo{person}{Thomas Sikor} {and} \bibinfo{person}{Peter Newell}.} \bibinfo{year}{2014}\natexlab{}.
\newblock \showarticletitle{Globalizing environmental justice?}
\newblock \bibinfo{journal}{\emph{Geoforum}}  \bibinfo{volume}{54} (\bibinfo{date}{July} \bibinfo{year}{2014}), \bibinfo{pages}{151--157}.
\newblock
\showISSN{00167185}
\href{https://doi.org/10.1016/j.geoforum.2014.04.009}{doi:\nolinkurl{10.1016/j.geoforum.2014.04.009}}


\bibitem[TAYLOR(2000)]%
        {TAYLOR_2000}
\bibfield{author}{\bibinfo{person}{DORCETA~E. TAYLOR}.} \bibinfo{year}{2000}\natexlab{}.
\newblock \showarticletitle{The rise of the Environmental Justice Paradigm}.
\newblock \bibinfo{journal}{\emph{American Behavioral Scientist}} \bibinfo{volume}{43}, \bibinfo{number}{4} (\bibinfo{date}{Jan} \bibinfo{year}{2000}), \bibinfo{pages}{508–580}.
\newblock
\href{https://doi.org/10.1177/0002764200043004003}{doi:\nolinkurl{10.1177/0002764200043004003}}


\bibitem[Temper(2019a)]%
        {temper_blocking_2019}
\bibfield{author}{\bibinfo{person}{Leah Temper}.} \bibinfo{year}{2019}\natexlab{a}.
\newblock \showarticletitle{Blocking pipelines, unsettling environmental justice: from rights of nature to responsibility to territory}.
\newblock \bibinfo{journal}{\emph{Local Environment}} \bibinfo{volume}{24}, \bibinfo{number}{2} (\bibinfo{date}{Feb.} \bibinfo{year}{2019}), \bibinfo{pages}{94--112}.
\newblock
\showISSN{1354-9839, 1469-6711}
\href{https://doi.org/10.1080/13549839.2018.1536698}{doi:\nolinkurl{10.1080/13549839.2018.1536698}}


\bibitem[Temper(2019b)]%
        {temper_blocking_2019-1}
\bibfield{author}{\bibinfo{person}{Leah Temper}.} \bibinfo{year}{2019}\natexlab{b}.
\newblock \showarticletitle{Blocking pipelines, unsettling environmental justice: from rights of nature to responsibility to territory}.
\newblock \bibinfo{journal}{\emph{Local Environment}} \bibinfo{volume}{24}, \bibinfo{number}{2} (\bibinfo{date}{Feb.} \bibinfo{year}{2019}), \bibinfo{pages}{94--112}.
\newblock
\showISSN{1354-9839, 1469-6711}
\href{https://doi.org/10.1080/13549839.2018.1536698}{doi:\nolinkurl{10.1080/13549839.2018.1536698}}


\bibitem[Temper et~al\mbox{.}(2018)]%
        {temper_perspective_2018}
\bibfield{author}{\bibinfo{person}{Leah Temper}, \bibinfo{person}{Mariana Walter}, \bibinfo{person}{Iokiñe Rodriguez}, \bibinfo{person}{Ashish Kothari}, {and} \bibinfo{person}{Ethemcan Turhan}.} \bibinfo{year}{2018}\natexlab{}.
\newblock \showarticletitle{A perspective on radical transformations to sustainability: resistances, movements and alternatives}.
\newblock \bibinfo{journal}{\emph{Sustainability Science}} \bibinfo{volume}{13}, \bibinfo{number}{3} (\bibinfo{date}{May} \bibinfo{year}{2018}), \bibinfo{pages}{747--764}.
\newblock
\showISSN{1862-4065, 1862-4057}
\href{https://doi.org/10.1007/s11625-018-0543-8}{doi:\nolinkurl{10.1007/s11625-018-0543-8}}


\bibitem[Tuck and Yang(2012)]%
        {tuck_decolonization_nodate}
\bibfield{author}{\bibinfo{person}{Eve Tuck} {and} \bibinfo{person}{K~Wayne Yang}.} \bibinfo{year}{2012}\natexlab{}.
\newblock \showarticletitle{Decolonization is not a metaphor}.
\newblock  (\bibinfo{year}{2012}).
\newblock


\bibitem[Ulloa(2017)]%
        {ulloa_perspectives_2017}
\bibfield{author}{\bibinfo{person}{Astrid Ulloa}.} \bibinfo{year}{2017}\natexlab{}.
\newblock \showarticletitle{Perspectives of {Environmental} {Justice} from {Indigenous} {Peoples} of {Latin} {America}: {A} {Relational} {Indigenous} {Environmental} {Justice}}.
\newblock \bibinfo{journal}{\emph{Environmental Justice}} \bibinfo{volume}{10}, \bibinfo{number}{6} (\bibinfo{date}{Dec.} \bibinfo{year}{2017}), \bibinfo{pages}{175--180}.
\newblock
\showISSN{1939-4071, 1937-5174}
\href{https://doi.org/10.1089/env.2017.0017}{doi:\nolinkurl{10.1089/env.2017.0017}}


\bibitem[United Church~of Christ(1987)]%
        {United_Church_of_Christ_1987}
\bibfield{author}{\bibinfo{person}{Commission for Racial~Justice United Church~of Christ}.} \bibinfo{year}{1987}\natexlab{}.
\newblock \bibinfo{title}{Toxic Wastes and Race In the United States}.
\newblock
\urldef\tempurl%
\url{https://www.nrc.gov/docs/ml1310/ml13109a339.pdf}
\showURL{%
\tempurl}


\bibitem[van~de Poel et~al\mbox{.}(2024)]%
        {van_de_poel_design_nodate}
\bibfield{author}{\bibinfo{person}{van~de Poel}, \bibinfo{person}{van Uffelen}, {and} \bibinfo{person}{Moreno Inglés}.} \bibinfo{year}{2024}\natexlab{}.
\newblock \showarticletitle{Design for {Justice}}.
\newblock  (\bibinfo{year}{2024}).
\newblock


\bibitem[Van~Horne et~al\mbox{.}(2023)]%
        {van_horne_applied_2023}
\bibfield{author}{\bibinfo{person}{Yoshira~Ornelas Van~Horne}, \bibinfo{person}{Cecilia~S. Alcala}, \bibinfo{person}{Richard~E. Peltier}, \bibinfo{person}{Penelope J.~E. Quintana}, \bibinfo{person}{Edmund Seto}, \bibinfo{person}{Melissa Gonzales}, \bibinfo{person}{Jill~E. Johnston}, \bibinfo{person}{Lupita~D. Montoya}, \bibinfo{person}{Lesliam Quirós-Alcalá}, {and} \bibinfo{person}{Paloma~I. Beamer}.} \bibinfo{year}{2023}\natexlab{}.
\newblock \showarticletitle{An applied environmental justice framework for exposure science}.
\newblock \bibinfo{journal}{\emph{Journal of Exposure Science \& Environmental Epidemiology}} \bibinfo{volume}{33}, \bibinfo{number}{1} (\bibinfo{date}{Jan.} \bibinfo{year}{2023}), \bibinfo{pages}{1--11}.
\newblock
\showISSN{1559-0631, 1559-064X}
\href{https://doi.org/10.1038/s41370-022-00422-z}{doi:\nolinkurl{10.1038/s41370-022-00422-z}}


\bibitem[Vera et~al\mbox{.}(2019)]%
        {vera_when_2019}
\bibfield{author}{\bibinfo{person}{Lourdes~A. Vera}, \bibinfo{person}{Dawn Walker}, \bibinfo{person}{Michelle Murphy}, \bibinfo{person}{Becky Mansfield}, \bibinfo{person}{Ladan~Mohamed Siad}, \bibinfo{person}{Jessica Ogden}, {and} \bibinfo{person}{{EDGI}}.} \bibinfo{year}{2019}\natexlab{}.
\newblock \showarticletitle{When data justice and environmental justice meet: formulating a response to extractive logic through environmental data justice}.
\newblock \bibinfo{journal}{\emph{Information, Communication \& Society}} \bibinfo{volume}{22}, \bibinfo{number}{7} (\bibinfo{date}{June} \bibinfo{year}{2019}), \bibinfo{pages}{1012--1028}.
\newblock
\showISSN{1369-118X, 1468-4462}
\href{https://doi.org/10.1080/1369118X.2019.1596293}{doi:\nolinkurl{10.1080/1369118X.2019.1596293}}


\bibitem[Whyte(2016)]%
        {Whyte_2016}
\bibfield{author}{\bibinfo{person}{Kyle Whyte}.} \bibinfo{year}{2016}\natexlab{}.
\newblock \showarticletitle{Indigenous experience, environmental justice and Settler colonialism}.
\newblock \bibinfo{journal}{\emph{SSRN Electronic Journal}} (\bibinfo{year}{2016}).
\newblock
\href{https://doi.org/10.2139/ssrn.2770058}{doi:\nolinkurl{10.2139/ssrn.2770058}}


\bibitem[Yanchapaxi and Murphy(2025)]%
        {yanchapaxi_indigenous_2025}
\bibfield{author}{\bibinfo{person}{María~Fernanda Yanchapaxi} {and} \bibinfo{person}{M. Murphy}.} \bibinfo{year}{2025}\natexlab{}.
\newblock \showarticletitle{Indigenous {Environmental} {Data} {Justice}: {Confronting} {Colonial} {Data} and {Activating} {Indigenous} {Sovereignty}}.
\newblock \bibinfo{journal}{\emph{Science, Technology, \& Human Values}} (\bibinfo{date}{June} \bibinfo{year}{2025}), \bibinfo{pages}{01622439251343837}.
\newblock
\showISSN{0162-2439, 1552-8251}
\href{https://doi.org/10.1177/01622439251343837}{doi:\nolinkurl{10.1177/01622439251343837}}


\end{thebibliography}

\end{document}